\documentclass[graybox]{svmult}

\usepackage{helvet}         
\usepackage{courier}        
\usepackage{type1cm}        
\usepackage{makeidx}         
\usepackage{graphicx}        
\usepackage{multicol}        
\usepackage[bottom]{footmisc}

\usepackage{amsmath}
\usepackage{amssymb}

\newcommand{\Real}{\mathop{\mathrm{Re}}\nolimits}
\newcommand{\Lin}{\mathop{\mathrm{Lin}}\nolimits}
\newcommand{\Tr}{\mathop{\mathrm{Tr}}\nolimits}
\newcommand{\ad}{\mathop{\mathrm{ad}}\nolimits}
\newcommand{\SL}{\mathrm{SL}}
\newcommand{\C}{\mathbb{C}}
\newcommand{\R}{\mathbb{R}}
\newcommand{\M}{\mathcal{M}}
\newcommand{\ii}{\mathrm{i}}
\newcommand{\Aut}{\mathop{\mathrm{Aut}}\nolimits}

\begin{document}

\title*{Possible alternative mechanism to SUSY: conservative extensions of the Poincar\'e group}
\titlerunning{Possible alternative mechanism to SUSY}
\author{Andr\'as L\'aszl\'o}
\institute{Andr\'as L\'aszl\'o \at Wigner Research Centre for Physics of the Hungarian Academy of Sciences, Konkoly-Thege M.\ u.\ 29-33.\ H-1121 Budapest Hungary, \email{laszlo.andras@wigner.mta.hu}}
\maketitle

\abstract{
A group theoretical mechanism is outlined, which can indecomposably 
extend the Poincar\'e group by the compact internal (gauge) symmetries 
at the price of allowing some nilpotent (or, more precisely: solvable) 
internal symmetries in addition. Due to the presence of this nilpotent 
part, the prohibitive argument of the well known Coleman-Mandula and McGlinn 
no-go theorems do not go through. 
In contrast to SUSY or extended SUSY, in our construction the symmetries 
extending the Poincar\'e group will be all internal, i.e.\ they do not 
act on the spacetime, merely on some internal degrees of freedom 
--- hence the name: \emph{conservative} extensions of the 
Poincar\'e group. Using the Levi decomposition and O'Raifeartaigh theorem, 
the general structure of all possible conservative extensions of the 
Poincar\'e group is outlined, and a concrete example group is 
presented with $\mathrm{U}(1)$ being the compact gauge group component. 
It is argued that such nilpotent internal symmetries may be inapparent 
symmetries of some more fundamental field variables, and therefore do not 
carry an ab initio contradiction with the present experimental understanding 
in particle physics. The construction is compared to (extended) SUSY, 
since SUSY is somewhat analogous to the proposed mechanism. 
It is pointed out, however, that the proposed mechanism is less irregular 
in comparison to SUSY, in certain aspects. The only exoticity needed in comparison 
to a traditional gauge theory setting is that the full group of internal 
symmetries is not purely compact, but is a semi-direct product of a nilpotent 
and of a compact part.
}

\section{Introduction}
\label{sec:Introduction}

In Lagrangian field theories it is well understood that larger amount of 
symmetries of the Lagrangian gives less room for variants of the theory. In 
particular, the larger amount of direct-indecomposable (unified) symmetries 
reduce the number of possible free coupling parameters. This phenomenon 
motivated the search for unified symmetries in field theory, meaning that a 
plausible direct-indecomposable symmetry group was being searched for, 
which contained the known symmetry groups as subgroups. When it comes to 
building relativistic field theories to be applied in particle physics, 
the known symmetry groups are the Poincar\'e group and the compact internal 
(gauge) symmetries of the Standard Model, commuting with each-other. Therefore, 
a rather plausible idea was to try to find a direct-indecomposable 
symmetry group, which contains Poincar\'e symmetries and compact internal 
symmetries, indecomposably. In 1964 it was realized by McGlinn 
\cite{mcglinn1964} that whenever the compact internal symmetries are 
semi-simple, this is group theoretically impossible. 
This motivated the work of O'Raifeartaigh in 1965 \cite{raifeartaigh1965} 
to try to understand all possible group extensions of the Poincar\'e 
symmetries. The pertinent O'Raifeartaigh theorem made it clear that the Lie 
group theoretical possibilities for a direct-indecomposable extension of 
the Poincar\'e group is rather limited. Historically, at the time of the 
publication of O'Raifeartaigh theorem, no constructive examples for the 
potentially allowed direct-indecomposable Poincar\'e group extensions were 
known. For instance supersymmetry (SUSY) was not known at the time, and 
the conformal Poincar\'e group, being a direct-indecomposable extension 
of the Poincar\'e group, was not in the physics folklore. Therefore, the 
potentially allowed Poincar\'e group extensions by means of the O'Raifeartaigh 
theorem were talked away by a littlebit handwaving physics arguments. Not much 
later, in 1967 the famous Coleman-Mandula theorem \cite{coleman1967} was 
published, stating that given some plausible assumptions, a unification of 
the Poincar\'e group with purely compact internal symmetries is not possible 
in the framework of quantum field theory. These attempts were historically 
reviewed in \cite{hegerfeldt1968}. A few years later, the famous paper 
of Wess and Zumino was published \cite{wess1974}, implicitly providing 
an example Lie group (the super-Poincar\'e group) which is an indecomposable 
extension of the Poincar\'e group, and thus providing an explicit example 
for one of the cases of O'Raifeartaigh theorem, allowing a 
direct-indecomposable extension of the Poincar\'e group. Motivated by this, 
Haag, Lopusza\'nski and Sohnius \cite{haag1975} generalized the 
Coleman-Mandula theorem also allowing for super-Poincar\'e transformations. 
Since that work, the so called super-Lie algebra view of those transformations 
is the most popular in the literature, making it less obvious to see the 
underlying ordinary Lie group structure of the super-Poincar\'e 
transformations, and their relations to O'Raifeartaigh theorem. In the recent 
years it was re-understood that there do exist also other direct-indecomposable 
extensions of the Poincar\'e group. A rather well-understood example is the 
conformal Poincar\'e group, being isomorphic to $\mathrm{SO}(2,4)$, but also 
others have been found \cite{nesti2010, lisi2010, chamseddine2016, percacci2008}, 
some of which can lead to field theories which may not be \emph{ab initio} pathological. 
They bypass the Coleman-Mandula theorem by weakening some of its assumptions, 
for instance allowing for symmetry breaking.

In this paper a newly found direct-indecomposable Poincar\'e group extension 
\cite{laszlo2016, laszlo2017} is discussed, which contains a Poincar\'e 
component, a compact internal group component, and a nilpotent internal group 
component. From the Lie group theoretical point of view, it resembles to 
the (extended) super-Poincar\'e group, since in its Levi decomposition its 
radical is a nilpotent Lie algebra. However, in contrast to SUSY, 
this group respects vector bundle structure of fields, i.e.\ all the 
non-Poincar\'e symmetries act spacetime pointwise on some internal degrees of 
freedom. This implies that symmetry breaking is not necessary in order to 
make this new symmetry concept to harmonize with a gauge-theory-like setting, 
where vector bundle structure of fundamental fields is essential to preserve. 
Hence, we call these constructions \emph{conservative} extensions of the 
Poincar\'e group.

The outline of the paper is as follows. In Section~\ref{sec:Levi} the 
general structure of Lie groups is recalled in the light of Levi decomposition 
theorem. In Section~\ref{sec:ORaifeartaigh} the O'Raifeartaigh classification theorem 
on Poincar\'e group extensions is recalled. In Section~\ref{sec:Conservative} 
the structure of conservative extensions of the Poincar\'e group is outlined. 
In Section~\ref{sec:Concrete} the Lie algebra of the concrete conservative 
Poincar\'e group extension defined in \cite{laszlo2016, laszlo2017} is presented.

\section{General structure of Lie groups: Levi decomposition}
\label{sec:Levi}

In every finite dimensional real Lie algebra, one has the Killing form, being a 
real valued bilinear form defined by the formula 
$x\cdot y:=\Tr\left(\ad_{x}\ad_{y}\right)$ for two elements $x,y$ of the Lie algebra. 
The Levi decomposition theorem \cite{onishchik1990, ise1991} states that 
the structure of a generic real finite dimensional connected and simply connected 
Lie group is as follows:
\begin{eqnarray}
 \underbrace{E}_{\mathrm{Lie\;group}} = \underbrace{R}_{\substack{\mathrm{degenerate\;directions\;of\;Killing\;form}\\\mathrm{(called\;to\;be\;the\;}radical\mathrm{)}}} \rtimes \underbrace{\big(L_{1} \times \dots \times L_{n}\big)}_{\substack{\mathrm{non-degenerate\;directions\;of\;Killing\;form}\\\mathrm{(called\;to\;be\;the\;}Levi\;factor\mathrm{)}}}
\end{eqnarray}
A subgroup spanned by the non-degenerate directions of the Killing form is 
called the \emph{Levi factor} or \emph{semisimple part}. It falls apart to 
direct product of subgroups which contain no proper normal subgroups, and 
are called the \emph{simple components}. The normal (invariant) subgroup 
spanned by the degenerate directions of the Killing form is called the 
\emph{radical} or \emph{solvable part}. The radical $R$ can also be 
equivalently characterized by the property that the Lie algebra $r$ of $R$ 
has terminating derived series. Namely, with the 
definition $r^{0}:=r$, $r^{k}:=\left[r^{k-1},r^{k-1}\right]$, there exists 
a finite $k$ such that $r^{k}=\{0\}$. A special case is when $R$ 
is said to be \emph{nilpotent}: in this case there exists a finite $k$ such that 
for all $x_{1},\dots,x_{k}\in r$ one has $\ad_{x_{1}}\dots\ad_{x_{k}}=0$. 
The extreme case is when $R$ is said to be \emph{abelian}: in this 
case for all $x\in r$ one has $\ad_{x}=0$.

Whenever also non-simply connected or non-connected Lie groups are considered, 
their generic structure can be slightly more complex:
\begin{eqnarray}
 \underbrace{E}_{\mathrm{Lie\;group}} = \Bigg( \Big( \underbrace{R}_{\mathrm{radical}} \rtimes \underbrace{\big(L_{1} \times \dots \times L_{n}\big)}_{\mathrm{Levi\;factor}} \Big) \big/ \underbrace{\mathcal{I}}_{\mathrm{discrete}} \Bigg) \;\rtimes\; \underbrace{\mathcal{J}}_{\mathrm{discrete}}
\end{eqnarray}
where $\mathcal{I}$ is some discrete normal subgroup of $R\rtimes\left(L_{1}\times\dots\times L_{n}\right)$ and 
$\mathcal{J}$ is some discrete subgroup of the outer automorphisms of the quotient group 
$\left(R\rtimes\left(L_{1}\times\dots\times L_{n}\right)\right)/\mathcal{I}$. 
It is not complicated to see that whenever a Lie group is injectively embedded 
into another, then its Lie algebra must be injectively embedded into the Lie 
algebra of the other. Thus, for studying necessary condition for injective 
embedding of Lie groups, one first needs to study the injective embeddings 
of Lie algebras, or equivalently, of connected and simply connected Lie groups. 
From now on, by Lie groups we shall always mean connected and simply connected 
ones, i.e.\ the universal covering groups.

Levi decomposition theorem can be illustrated with the Poincar\'e group:
\begin{eqnarray}
 \underbrace{\mathcal{P}}_{\mathrm{Poincar\acute{e}\;group}} = \underbrace{\mathcal{T}}_{\mathrm{translations\;(radical)}} \rtimes \underbrace{\mathcal{L}}_{\mathrm{Lorentz\;group\;(Levi\;factor)}}
\end{eqnarray}

\section{A classification of Poincar\'e group extensions}
\label{sec:ORaifeartaigh}

A classification scheme of Poincar\'e group extensions was outlined 
by O'Rai\-fear\-taigh \cite{raifeartaigh1965}, using the Levi decomposition 
theorem. It is based on the simple observation that when injectively embedding 
a finite dimensional real Lie algebra into another, then the Levi factor 
of the smaller Lie algebra cannot intersect with the radical of the larger one. 
This implies the following disjoint possibilities for a 
connected and simply connected extension 
$E=R\rtimes\left(L_{1}\times\dots\times L_{n}\right)$ of the Poincar\'e symmetries 
$\mathcal{P}=\mathcal{T}\rtimes\mathcal{L}$.
\vspace*{-1mm}
\begin{itemize}
\item[ A] One has $E=\mathcal{P}\times\{\mathrm{some\;other\;Lie\;group}\}$, i.e. no unification occurs.
\item[ B] One has not A and $\mathcal{T}\subset R$ and $\mathcal{L}\subset L_{1}$, meaning 
that the translations $\mathcal{T}$ are injected into the radical $R$ and the homogeneous 
Lorentz group $\mathcal{L}$ is injected into one of the simple components $L_{1}$ of $E$.
\item[ C] One has $(\mathcal{T}{\rtimes}\mathcal{L})\subset L_{1}$, i.e.\ the 
entire Poincar\'e group is injected into one of the simple components $L_{1}$ of $E$.
\end{itemize}

Examples for case B are detailed in \cite{laszlo2017}, namely the super-Poincar\'e group or the extended super-Poincar\'e 
group \cite{wess1974, salam1974, ferrara1974}, as well as the extensions of 
the Poincar\'e group proposed by us \cite{laszlo2017}. Example for case C is the conformal Poincar\'e group, 
being isomorphic to $\mathrm{SO}(2,4)$. However, also more complicated examples 
are being constructed \cite{nesti2010, lisi2010, chamseddine2016} in the literature.

Knowing O'Raifeartaigh theorem, the argument of Coleman-Mandula theorem 
in case of a finite dimensional Poincar\'e group extension can be greatly 
simplified. First, Coleman-Mandula assumes implicitly that symmetry breaking 
is not present, which excludes case C. Secondly, it implicitly assumes 
that one has a positive definite invariant scalar product on the 
non-Poincar\'e directions of the Lie algebra, which excludes case B 
(along with SUSY, for instance). In case of SUSY or our Poincar\'e group 
extensions, the pertinent invariant scalar product is merely positive \emph{semi}definite, 
which provides a backdoor to the otherwise prohibitive argument.

\section{Conservative extensions of the Poincar\'e group}
\label{sec:Conservative}

As outlined in \cite{laszlo2017}, the super-Poincar\'e group or extended 
super-Poincar\'e group cannot be considered as a vector bundle automorphism 
group with the spacetime being the base manifold. This implies that in a 
supersymmetric model a heavy symmetry breaking needs to be introduced in order to 
recover a gauge-theory-like setting, so characteristic to the Standard Model. 
Also in \cite{laszlo2017} the question is asked: what are those finite dimensional 
direct-indecomposable extensions $E$ of the Poincar\'e group $\mathcal{P}=\mathcal{T}\rtimes\mathcal{L}$, 
which respect the vector bundle structure of fundamental fields as well as 
the Lorentz metric of the spacetime? Technically, this means that one has 
$E=\mathcal{T}\rtimes\{\mathrm{some\;pointwise\;acting\;symmetries}\}\,$ with 
a surjective homomorphism $\,\{\mathrm{some\;pointwise\;acting\;symmetries}\}\rightarrow\mathcal{L}\,$ 
onto the Lorentz group. The answer \cite{laszlo2017} is a simple consequence of the Levi decomposition / O'Raifeartaigh 
theorem and of the definition of semidirect product:
\begin{eqnarray}
 E = \big(\underbrace{\mathcal{T}}_{\mathrm{translations}} \times \underbrace{\mathcal{N}}_{\substack{\mathrm{solvable}\\\mathrm{internal\;symmetries}}}\big) \rtimes \big(\underbrace{\mathcal{G}_{1}{\times}{\dots}{\times}\mathcal{G}_{m}}_{\substack{\mathrm{semisimple}\\\mathrm{internal\;symmetries}}} \times \underbrace{\mathcal{L}}_{\substack{\mathrm{Lorentz}\\\mathrm{symmetries}}}\big)
\end{eqnarray}
must hold, where the semisimple internal symmetries $\mathcal{G}_{1}{\times}{\dots}{\times}\mathcal{G}_{m}$ 
commute with the translations $\mathcal{T}$, the Lorentz symmetries $\mathcal{L}$ have the 
canonical adjoint action on the translations $\mathcal{T}$, but the 
invariant subgroup of solvable internal symmetries $\mathcal{N}$ does not commute 
with the Lorentz symmetries nor with the semisimple internal symmetries. If one requires in addition that there 
exists a positive \emph{semi}definite invariant bilinear form on the Lie algebra 
of the non-Poincar\'e symmetries, then it also follows that 
$\mathcal{G}_{1}{\times}{\dots}{\times}\mathcal{G}_{m}$ is compact. (Such a requirement 
is motivated by the positive energy condition for gauge fields.) With 
this requirement, the full internal symmetry group of such a Poincar\'e group 
extension shall have the structure $\{\mathrm{solvable}\}\rtimes\{\mathrm{compact}\}$. 
These kind of Poincar\'e group extensions we named \emph{conservative} extensions, and 
are seen to have a number of rather favorable properties \cite{laszlo2017}: 
they are direct-indecomposable, preserve causal structure of the spacetime, 
preserve vector bundle structure of fundamental fields, 
obey positive energy condition etc. Ideally, one could look for such a setting 
in which case the group of compact internal symmetries is identical to the 
Standard Model gauge group $\mathrm{U}(1){\times}\mathrm{SU}(2){\times}\mathrm{SU}(3)$.

It is not difficult to see that conservative extensions of the Poincar\'e group 
do exist, i.e.\ that our definition is not empty. Take, for instance, the complexified 
Schr\"odinger Lie group, which is isomorphic to $\mathrm{H}_{3}(\C)\rtimes \SL(2,\C)$. 
Here $H_{3}(\C)$ denotes the complexified Heisenberg Lie group with three generators, 
being the lowest dimensional complex non-abelian nilpotent Lie group. Clearly, from 
this semi-direct product there exists a homomorphism onto $\SL(2,\C)$ and therefore also onto 
the homogeneous Lorentz group $\mathcal{L}$, which acts canonically on the group of spacetime 
translations $\mathcal{T}$ in its adjoint representation. With these subgroup actions, the group 
$\left(\mathcal{T}\times \mathrm{H}_{3}(\C)\right)\rtimes \mathcal{L}$ is uniquely well-defined 
and is direct-indecomposable. (Note that from the Lie algebra point of view, one 
has $\SL(2,\C)\equiv\mathcal{L}$). This provides the simplest conservative extension of the Poincar\'e group, 
and the non-Poincar\'e symmetries span a nilpotent Lie group $\mathrm{H}_{3}(\C)$, being part of the radical.

An other example is constructed in \cite{laszlo2016, laszlo2017}, which is expected to be 
more interesting for physics. It 
contains a Poincar\'e component, a compact internal group component ($\mathrm{U}(1)$ in the example), 
and unavoidably a nilpotent internal group component. In particular, 
it has the group structure $\left(\mathcal{T}\times N\right)\rtimes \left(\mathrm{U}(1)\times\mathcal{L}\right)$, 
where $N$ is a 20 dimensional real nilpotent Lie group, the Lorentz group $\mathcal{L}$ 
acts with the canonical adjoint action on the translations $\mathcal{T}$, 
and both the compact $\mathrm{U}(1)$ component and the Lorentz group component 
$\mathcal{L}$ has non-vanishing adjoint action on $N$, which provides 
the direct-indecomposability. Clearly, it is essential in the construction that 
the radical $\mathcal{T}$ of the Poincar\'e group is extended by $N$, 
without which such a direct-indecomposability is not possible 
according to O'Raifeartaigh theorem. Also note, that the construction 
resembles to (extended) super-Poincar\'e group as outlined in \cite{laszlo2017}, 
with the important difference that in case of the (extended) super-Poincar\'e 
group the translations are direct-indecomposably part of the nilpotent 
symmetries, called to be the group of supertranslations, forming a 
direct-indecomposable two-step nilpotent Lie group. In case of our construction, 
however, the translations are direct-decomposable from other symmetries within the 
radical, which makes it a conservative extension of the Poincar\'e group, in 
contrast to (extended) SUSY. It is also an important piece of information 
that the concrete conservative extension of the Poincar\'e group proposed 
in \cite{laszlo2016, laszlo2017} can be shown to have faithful unitary 
representations on some separable complex Hilbert space.

An important feature of the conservative extensions of the Poincar\'e group $\mathcal{P}$ 
is that there exists a homomorphism:
\begin{eqnarray}
\underbrace{ \underbrace{\mathcal{N}}_{\substack{\mathrm{solvable}\\\mathrm{internal\;symmetries}}} \rtimes \bigg( \underbrace{\mathcal{G}_{1}{\times}{\dots}{\times}\mathcal{G}_{m}}_{\substack{\mathrm{compact}\\\mathrm{internal\;symmetries}}} \times \underbrace{\mathcal{P}}_{\substack{\mathrm{Poincar\acute{e}}\\\mathrm{symmetries}}} \bigg) }_{\substack{\mathrm{direct-indecomposable\;conservative\;extension\;of\;the\;Poincar\acute{e}\;group,}\\\mathrm{acting\;on\;fundamental\;field\;degrees\;of\;freedom}}} \qquad\qquad\qquad\qquad \cr
 \cr
 \cr
 \qquad\qquad \longrightarrow 
\underbrace{ \underbrace{\mathcal{G}_{1}{\times}{\dots}{\times}\mathcal{G}_{m}}_{\substack{\mathrm{compact}\\\mathrm{internal\;symmetries}}} \times \underbrace{\mathcal{P}}_{\substack{\mathrm{Poincar\acute{e}}\\\mathrm{symmetries}}} }_{\substack{\mathrm{observed\;direct-decomposable\;symmetries,}\\\mathrm{acting\;on\;some\;derived\;field\;quantities}\\\mathrm{which\;are\;function\;of\;fundamental\;degrees\;of\;freedom}}} 
\end{eqnarray}
and potentially can explain a Standard Model-like gauge theory setting from 
a direct-indecomposable fundamental symmetry, without a breaking of it.

\section{Commutation relations of the concrete example}
\label{sec:Concrete}

In this section the commutation relations of the generators of the Lie algebra 
of our concrete example group \cite{laszlo2016, laszlo2017} is outlined. 
The pertinent direct-indecomposable conservative extension of the Poincar\'e 
group is the automorphism group of some finite dimensional unital associative 
algebra valued classical fields over the four dimensional spacetime. Similar algebra 
valued field construction was tried by Anco and Wald in the end of '80-s 
\cite{anco1989}, but they could not achieve the goal of 
direct-indecomposability due to the too simple structure of the algebra 
of fields which they applied.

In the followings $S$ shall denote a complex two-dimensional vector space 
(``spinor space''), and $S^{*}$, $\bar{S}$, $\bar{S}^{*}$ shall denote its 
dual, complex conjugate, complex conjugate dual vector space, respectively. 
Let us consider the complex unital associative algebra 
$\Lambda(\bar{S}^{*})\otimes\Lambda(S^{*})$, where $\Lambda()$ denotes 
exterior algebra formation. Observe that this algebra also 
has an antilinear involution defined by the complex conjugation, which is 
compatible with the algebraic product in the sense that $\overline{x\,y}=\bar{x}\,\bar{y}$ 
holds for any two algebra elements $x,y$. We shall call a finite dimensional 
complex unital associative algebra $A$ together with an antilinear involution 
$(\cdot)^{+}$ obeying $(x\,y)^{+}=x^{+}\,y^{+}$ a \emph{spin algebra} whenever 
the pair $\left(A,\,(\cdot)^{+}\right)$ is isomorphic to 
$\left(\Lambda(\bar{S}^{*})\otimes\Lambda(S^{*}),\,\overline{(\cdot)}\right)$. 
The antilinear involution $(\cdot)^{+}$ (or, $\overline{(\cdot)}$) shall be referred 
to as \emph{charge conjugation}. Thus, a spin algebra $A$ is (not naturally) 
isomorphic to the concrete spin algebra $\Lambda(\bar{S}^{*})\otimes\Lambda(S^{*})$ 
with spinorial realization. In the followings, we shall often use a 
representation $A\cong \Lambda(\bar{S}^{*})\otimes\Lambda(S^{*})$ so that 
the simple formalism of traditional two-spinor calculus can be used.

For the sake of simplicity, we shall give our construction in the flat 
spacetime limit. Let $\M$ denote a four real dimensional affine space, 
modeling a (flat) spacetime manifold, and let $T$ be its underlying vector 
space (``tangent space''). Take the trivial vector bundle $A(\M):=\M\times A$. 
Our direct-indecomposable conservative Poincar\'e group extension containing 
also $\mathrm{U}(1)$ shall be nothing but the automorphism group of the 
algebra of the sections of the $A(\M)$, i.e.\ of the spin algebra valued fields 
\cite{laszlo2016, laszlo2017}. In the followings Penrose abstract indices shall 
be used for the spacetime degrees of freedom and for the spinor degrees of 
freedom, as usual in the General Relativity literature 
\cite{penrose1984, wald1984}. The symbol $\nabla_{a}$ 
shall denote the affine covariant derivation of the affine space $\M$. 
Also, given a point $o$ (``origin'') of $\M$, the symbol $X_{o}$ shall 
denote the vectorization map against $o$, which is the vector field 
$X_{o}:\;\M\rightarrow T,\;x\mapsto ({x}{-}{o})$. Let in the spinorial 
representation $\sigma_{a}^{AA'}$ denote the usual Infeld-Van der Waerden 
symbol, also called Pauli injection, or soldering form. It is some preferred injective linear map 
$T\rightarrow \Real\left(\bar{S}\otimes S\right)$, and is shown in \cite{laszlo2016, laszlo2017} 
to be $\Aut(A)$-invariant. Its inverse map is denoted 
by $\sigma^{a}_{AA'}$. Let $\omega_{[A'B'][CD]}$ be a positive maximal form 
from $A$. Then, it is well-known that 
$g(\sigma,\omega)_{ab}:=\sigma_{a}^{AA'}\sigma_{b}^{BB'}\omega_{[A'B'][AB]}$ is 
a Lorentz signature metric on $T$, and its inverse metric is denoted 
by $g(\sigma,\omega)^{ab}$. The symbol 
$\Sigma(\sigma)_{a}{}^{b}{}_{B}{}^{A}:=\ii\left(\sigma_{a}^{AC'}\sigma^{b}_{BC'}-g(\sigma,\omega)^{cb}g(\sigma,\omega)_{da}\sigma_{c}^{AC'}\sigma^{d}_{BC'}\right)$ 
is called the spin tensor in the literature, and can be considered as the 
generators of the $\SL(2,\C)$ group, as it is well-known. It can uniquely 
act on the full mixed tensor algebra of $S$, $S^{*}$, $\bar{S}$, $\bar{S}^{*}$ 
by requiring vanishing action on scalars, commutativity with duality form, 
realness of $\ii\Sigma(\sigma)_{a}{}^{b}$, and Leibniz rule over tensor product. 
Given a concrete spinorial representation 
$A\equiv\Lambda(\bar{S}^{*})\otimes\Lambda(S^{*})$, thus the spin tensor can be 
uniquely extended to $A$ as an algebra derivation valued tensor 
$\Sigma(\sigma)_{a}{}^{b}$, and it shall have vanishing action on scalars, 
shall obey Leibniz rule against algebra multiplication of $A$, and shall have realness of 
$\ii\Sigma(\sigma)_{a}{}^{b}$ against the charge conjugation within $A$. 
The spin tensor $\Sigma(\sigma)_{a}{}^{b}$, however, is not invariant to the full action of 
$\Aut(A)$: the nilpotent normal subgroup within $\Aut(A)$ which do not preserve 
the subspaces $\Lambda_{\bar{p}q}:=\mathop{\wedge}^{p}\bar{S}^{*}\mathop{\otimes}\mathop{\wedge}^{q}S^{*}$ 
of pure $p,q$-forms do not preserve $\Sigma(\sigma)_{a}{}^{b}$. That is, the 
definition of $\Sigma(\sigma)_{a}{}^{b}$ is relative to a concrete spinorial 
representation $A\equiv\Lambda(\bar{S}^{*})\otimes\Lambda(S^{*})$, which is also 
not preserved by the pertinent nilpotent normal subgroup.

Introduce the differential operators 
${J_{o}}^{ab} := \left({X_{o}}^{a}\ii\nabla^{b}-{X_{o}}^{b}\ii\nabla^{a}\right) + \frac{1}{2}\Sigma^{ab}$ and 
$P_{a}:=\ii\nabla_{a}$
over the sections of the spin algebra bundle $A(\M)$, i.e.\ over the 
spin algebra valued fields. They are called the $o$-angular momentum and 
momentum operators, respectively, and are known to provide a faithful representation 
of the Poincar\'e Lie algebra in the Lie algebra of differential operators on 
the sections of $A(\M)$. Given a concrete spinorial 
representation $A\equiv\Lambda(\bar{S}^{*})\otimes\Lambda(S^{*})$, for each 
complex number $c$ introduce the unique algebra derivation operator which 
acts as $\zeta_{c}(\bar{\xi}_{A'}):=c\,\bar{\xi}_{A'}$ for all $\bar{\xi}_{A'}\in\Lambda_{\bar{1}0}\equiv\mathop{\wedge}^{1}\bar{S}^{*}\mathop{\otimes}\mathop{\wedge}^{0}S^{*}$. 
By construction, the map $\ii\,\varphi \mapsto \zeta_{\ii\,\varphi}$ $\;$ ($\varphi\in\R$) 
provides a faithful representation of the Lie algebra of the $\mathrm{U}(1)$ 
group on the algebra derivations of the spin algebra $A$, and thus on the algebra 
derivations of the spin algebra valued fields. Similarly to the spin tensor 
$\Sigma^{ab}$, the definition of the operator $\zeta$ depends on a concrete 
chosen spinorial representation $A\equiv\Lambda(\bar{S}^{*})\otimes\Lambda(S^{*})$. 
By construction, the operators $P_{a}$, ${J_{o}}_{\,ab}$, $\zeta$ provide a 
faithful representation of the Lie algebra of $\mathcal{P}\times\mathrm{U}(1)$.

The direct-indecomposable unification of $\mathcal{P}$ and of $\mathrm{U}(1)$ 
shall happen because $\Aut(A)$ has a nilpotent normal subgroup on which 
both $\mathcal{P}$ and $\mathrm{U}(1)$ has nonvanishing adjoint action. The 
generators of this nilpotent normal subgroup shall be detailed as follows. 
Take a concrete chosen spinorial representation 
$A\equiv\Lambda(\bar{S}^{*})\otimes\Lambda(S^{*})$. 
Take any element $\beta\in\Real\left(\Lambda_{\bar{1}2}\otimes\Lambda_{\bar{1}0}^{*}\oplus\Lambda_{\bar{2}1}\otimes\Lambda_{\bar{0}1}^{*}\right)\subset\Real\left(\Lin(A)\right)$. 
Such an element, in the spinorial notation, can be represented as 
$\left(\beta_{B'[CD]}{}^{A'},\,\bar{\beta}_{B[C'D']}{}^{A}\right)$, uniquely 
determined by the spinor tensor $\beta_{B'[CD]}{}^{A'}$. Such an element $\beta$ 
defines a $\Lambda_{\bar{1}0}\rightarrow\Lambda_{\bar{1}2}$ linear operator via 
the formula $\bar{\xi}_{A'}\mapsto\beta_{B'[CD]}{}^{A'}\bar{\xi}_{A'}$. 
Direct verification shows that this can be uniquely extended as an algebra 
derivation operator $\nu_{\beta}$ of $A$, via requiring vanishing on scalars 
$\Lambda_{\bar{0}0}$, realness, and Leibniz rule. 
Also, for all elements $a\in\Real\left(A\right)$, the linear map $\ad_{a}:A\rightarrow A$ 
is an algebra derivation of $A$, called inner derivation. They can be uniquely 
parameterized by real elements not in the center of $A$, i.e.\ with elements 
$a\in\Real\left(\Lambda_{\bar{1}0}\oplus\Lambda_{\bar{0}1}\oplus\Lambda_{\bar{1}1}\oplus\Lambda_{\bar{2}1}\oplus\Lambda_{\bar{1}2}\right)$.

Let 
$\beta,\beta'\in\Real\left(\Lambda_{\bar{1}2}\otimes\Lambda_{\bar{1}0}^{*}\oplus\Lambda_{\bar{2}1}\otimes\Lambda_{\bar{0}1}^{*}\right)\subset\Real\left(\Lin(A)\right)$ and 
take the elements $a,a'\in\Real\left(\Lambda_{\bar{1}0}\oplus\Lambda_{\bar{0}1}\oplus\Lambda_{\bar{1}1}\oplus\Lambda_{\bar{2}1}\oplus\Lambda_{\bar{1}2}\right)$ 
and $\varphi,\varphi'\in\R$, regarded as constant fields over the spacetime manifold $\M$. 
Then the relations
\begin{eqnarray}
\left[\ad_{a},\,\ad_{a'}\right] & = & \ad_{[a,a']}, \cr
\left[\ad_{a},\,\nu_{\beta'}\right] & = & -\ad_{\nu_{\beta'}(a)}, \cr
\left[\ad_{a},\,\zeta_{\ii\varphi'}\right] & = & -\ad_{\zeta_{\ii\varphi'}(a)}, \cr
\left[\ad_{a},\,{J_{o}}_{\,cd}\right] & = & -\ad_{{J_{o}}_{\,cd}(a)}, \cr
\left[\ad_{a},\,P_{c}\right] & = & 0, \cr
\left[\nu_{\beta},\,\nu_{\beta'}\right] & = & 0, \cr
\left[\nu_{\beta},\,\zeta_{\ii\varphi'}\right] & = & -\nu_{[\zeta_{\ii\varphi'},\beta]}, \cr
\left[\nu_{\beta},\,{J_{o}}_{\,cd}\right] & = & -\nu_{[{J_{o}}_{\,cd},\beta]}, \cr
\left[\nu_{\beta},\,P_{c}\right] & = & 0, \cr
\left[\zeta_{\ii\varphi},\,\zeta_{\ii\varphi'}\right] & = & 0, \cr
\left[\zeta_{\ii\varphi},\,{J_{o}}_{\,cd}\right] & = & 0, \cr
\left[\zeta_{\ii\varphi},\,P_{c}\right] & = & 0, \cr
\left[{J_{o}}_{\,cd},\,{J_{o}}_{\,ef}\right] & = & \ii\,g_{de}\,{J_{o}}_{\,cf} - \ii\,g_{ce}\,{J_{o}}_{\,df} + \ii\,g_{cf}\,{J_{o}}_{\,de} - \ii\,g_{df}\,{J_{o}}_{\,ce}, \cr
\left[{J_{o}}_{\,cd},\,P_{e}\right] & = & \ii\,g_{de}\,P_{c} - \ii\,g_{ce}\,P_{d}, \cr
\left[P_{c},\,P_{d}\right] & = & 0
\label{eq:lierel}
\end{eqnarray}
are seen to hold, where the operators $\ad_{a}$, $\nu_{\beta}$, $\zeta_{\ii\varphi}$, ${J_{o}}_{\,cd}$, $P_{e}$ 
are regarded as acting on the smooth sections of $A(\M)$, i.e.\ on spin algebra 
valued fields. These operators are algebra derivation valued on the algebra of 
smooth sections of $A(\M)$, where in a concrete spinor representation 
$A\equiv\Lambda(\bar{S}^{*})\otimes\Lambda(S^{*})$, these 
fields can be regarded as a 9-tuple of spinor tensor fields
\begin{minipage}{\textwidth}
\begin{eqnarray}
\Big(\varphi,\; \xi_{(+)\,A'},\; \xi_{(-)\,A'},\; \epsilon_{(+)\,[B'C']},\; v_{DD'},\; \epsilon_{(-)\,[BC]}, \qquad\qquad\qquad\qquad \cr
 \qquad\qquad\qquad\qquad \chi_{(+)\,[B'C']A},\; \chi_{(-)\,A'[BC]},\; \omega_{[A'B'][CD]}\Big)
\end{eqnarray}
\end{minipage}
in the usual spinor index notation. 
The symmetry generators in Eq.(\ref{eq:lierel}) respect the vector bundle structure of $A(\M)$, 
the spin algebra structure of the fibers of $A(\M)$, as well as the soldering 
form $\sigma_{a}^{AA'}$ viewed as a $T^{*}\otimes\Real\left(\Lambda_{\bar{1}1}^{*}\right)$ 
valued constant field over the affine space $\M$. They also happen to preserve 
the constant maximal forms $\omega_{[A'B'][CD]}$, i.e.\ constant sections of 
value in $\Lambda_{\bar{2}2}\equiv\mathop{\wedge}^{2}\bar{S}^{*}\mathop{\otimes}\mathop{\wedge}^{2}S^{*}$. 
If an additional generator, i.e.\ the operator $\rho\mapsto\zeta_{\rho}$ $\;$ ($\rho\in\R$) 
is also included among the Lie algebra generators of Eq.(\ref{eq:lierel}), then 
also the generators of the constant Weyl (conformal) rescalings of the flat 
spacetime metric $g_{ab}$ is included in the Lie algebra, and in that case 
the maximal forms are not preserved, but acted on with the Weyl rescalings.

\begin{acknowledgement}
The author would like to thank to the organizers of QTS10. Special thanks to 
prof.\ Evgeny Ivanov for discussions on a simplified example group for 
extended Poincar\'e symmetries, based on the complexified Schr\"odinger group. 
This work was supported in part by the J\'anos Bolyai Research Scholarship 
of the Hungarian Academy of Sciences. Financial coverage for the 
participation of the author on QTS10 was provided in part via the 
Momentum (`Lend\"ulet') program of the Hungarian Academy of Sciences under 
grant number LP2013-60, special thanks to Dezs\H{o} Varga for that. Moreover, 
significant part of the conference participation was covered by the travel 
grant of the Wigner Research Centre for Physics of the Hungarian Academy 
of Sciences, which is also greatly acknowledged.
\end{acknowledgement}

\end{document}